\documentstyle[aps,prb,multicol,epsf]{revtex} 
\begin{document}
\draft
\title{Periodic orbit theory for realistic cluster potentials:\\ 
       The leptodermous expansion}
\author{Erik Koch}
\address{Max-Planck-Institut f\"ur Festk\"orperforschung, D-70569 Stuttgart}
\date{\today}

\maketitle 
\begin{abstract}
The formation of supershells observed in large metal clusters can be 
qualitatively understood from a periodic-orbit-expansion for a spherical 
cavity. To describe the changes in the supershell structure for different 
materials, one has, however, to go beyond that simple model. We show how 
periodic-orbit-expansions for realistic cluster potentials can be derived 
by expanding {\em only} the classical radial action around the limiting case 
of a spherical potential well. We give analytical results for the leptodermous 
expansion of Woods-Saxon potentials and show that it describes the shift of 
the supershells as the surface of a cluster potential gets softer. As a 
byproduct of our work, we find that the electronic shell and supershell 
structure is not affected by a lattice contraction, which might be present 
in small clusters.
\end{abstract}

\begin{multicols}{2}
\section{Introduction}

One of the most surprising aspects of the physics of metal clusters is the
supershell structure observed in mass-abundance spectra.\cite{cpl91,nature91,%
Brechignac93,Broyer93} This feature can be traced back to a beating pattern
in the density of states for typical cluster potentials.\cite{nishioka90}.
The conceptual framework for understanding how this quantum interference comes
about is provided by periodic-orbit-theory.\cite{BaBlo3,gutzwiller70} 
The elegance of this approach rests on the fact that the 
periodic-orbit-expansion (POE) is known analytically for the 
spherical cavity. For this model potential it was found that the most important
contributions to the oscillating part of the density of states stem from the 
two shortest planar periodic orbits: the triangular and the square 
orbits. Since these contributions oscillate with similar frequencies, their
interference gives rise to a beating pattern, hence supershells.

Although a spherical potential well is a good first approximation to a cluster
potential, this model clearly cannot account for the changes in the electronic 
shell and supershell structure observed for clusters made of different 
materials. 
It is therefore desirable to understand how the periodic-orbit-expansion
is modified as one considers more realistic model potentials. A straightforward
approach for doing so is to solve the action integrals, which lie at the heart
of periodic-orbit-theory, numerically. That way, however, most of the elegance
and power of the periodic-orbit-expansion is lost. Analytical expression,
on the other hand, may well reveal the relevant parameters determining 
the supershell structure, and provide insight into how it changes as the 
cluster potential is varied. They should prove especially helpful in the
search for better self-consistent models, which properly describe the 
experimental data. 

While the simple spherical, homogeneous jellium 
model\cite{Beck,ekardt84} works quite well for the alkali clusters, it fails
to describe the supershell structure observed in Ga$_N$.\cite{Broyer93}
Attempts to improve the situation include, e.g.\ 
the introduction of smooth jellium-profiles,\cite{Lerme93b} 
the inclusion of pseudopotentials,\cite{Lerme94,Lerme95a,Lerme96} or 
the consideration of surface roughness.\cite{Lerme95b}

The present work arises from the desire to understand the simple spherical,
homogeneous jellium model.  An analysis of the density dependence of the 
electronic supershells in jellium clusters showed that the supershells are 
shifted as the potential at the cluster surface becomes softer.\cite{jelly} 
It has been demonstrated that this shift can be understood in the framework 
of a periodic-orbit-expansion for typical self-consistent cluster 
potentials.\cite{dissl} The purpose of the present paper is to give a
derivation of the leptodermous expansion, which requires the linearization of
{\em only the radial action}. We furthermore analyze the validity of the 
approximations involved and show comparisons with quantum mechanical 
calculations.

To set the stage for the semiclassical treatment of electronic supershells,
Sec.~\ref{shellcorr} gives a review of the shell correction 
methods.\cite{Strutinsky,Brack72,HarrisArgument} These
methods establish a systematic relation between self-consistent calculations
and one-electron calculations for suitable model potentials. We stress the 
fact that it is decisive to choose families of potentials that vary smoothly 
with cluster size, to describe the electronic shell structure properly.

Sec.~\ref{POE} is devoted to periodic-orbit-expansions. We sketch
the derivation of the POE for the oscillating part $\tilde{\rho}$ of the 
density of states using the path integral formalism along the lines given 
by Gutzwiller.\cite{gutzwiller70} Special attention is paid to the rate of 
convergence of the sum over classical periodic orbits. Since the shell and 
supershell structure observed in the mass spectra of metal clusters is not 
directly linked to $\tilde{\rho}$ but rather the variations $\tilde{E}$ in 
{\em total energy}, we proceed to derive a periodic-orbit-expansion for 
$\tilde{E}$. We  find that the latter expansion converges much more rapidly 
than that for the density of states, hence making any artificial smoothing 
of the spectrum, commonly introduced to lessen the contribution of the longer 
orbits to $\tilde{\rho}$,\cite{BaBlo3,nishioka90,bennemannprl} superfluous. 
To assess the validity of the expression for $\tilde{E}$ we check the 
truncated POE against the quantum-mechanical result. This comparison shows that
in the size range, which seems experimentally accessible, $\tilde{E}$ is well
described by a truncated POE, taking only triangular and square orbits into
account. This justifies the common practice of truncating the POE after the
two shortest {\em planar} orbits. 

As an immediate application of the periodic-orbit-expansion for $\tilde{E}$ 
we show in Sec.~\ref{lattcontr} that the electronic shell structure
is virtually not affected by any lattice contraction.
Such an increase in density for small clusters was suggested by EXAFS 
analyses\cite{EXAFS1,EXAFS2} and seems also plausible from the viewpoint
of continuum mechanics, which implies that the surface tension should lead to 
a compression of the smaller clusters.
We show that the major effect of a lattice contraction on $\tilde{E}$ is a
change of the amplitude for small clusters. The position of the shell minima
and the supershell structure is, however, not noticeably changed, even for
unrealistically large contractions. 

In Sec.~\ref{lepto} we show how to extend the 
periodic-orbit-expansion of $\tilde{E}$ to more realistic potentials.
We start from the observation that the surface width $a$ of the cluster 
potential is an important parameter determining the supershell 
structure.\cite{jelly} The basic idea is then to expand the action integrals
entering the POE around the analytically known results for a potential well.
It turns out that the actions can be very
well approximated by linear functions in the surface parameter $a/R_0$, where
$R_0$ is the radius of the cluster. Thus a finite surface width leaves the
frequencies in the POE unchanged and, to first order, only introduces phase
shifts. Taking also the change of the Fermi energy into account, we can
understand the changes in the electronic shells and supershells 
introduced by a soft potential surface. 

The technical details of the leptodermous expansion for Woods-Saxon potentials
are described in the appendix.

To simplify the notation we set $\hbar^2/2m$ to unity, 
i.e.\ we give lengths in Bohr-radii ($a_0$) and energies in Rydberg. 

\section{Shell Correction Methods}
\label{shellcorr}

The total energy $E(N)$ of clusters having $N$ valence electrons can be split
into a smooth and an oscillating part:
\begin{equation}
 E(N) = \bar{E}(N) + \tilde{E}(N) .
\end{equation}
The smooth part describes the overall change in energy as the cluster size 
increases and is given by a liquid drop expansion\cite{PerdewLiquidDrop,%
PerdewClust} 
\begin{equation}
 \bar{E}(N) = a_1\,N + a_2\,N^{2/3} + a_3\,N^{1/3} + \cdots.
\end{equation}
The oscillating part is responsible for the shell structure. 

The idea of shell correction methods is to give a prescription for determining 
$\tilde{E}(N)$ from a one-particle calculation. These methods were pioneered by
Strutinsky, who showed how the oscillating part of the total energy resulting 
from a Hartree-Fock calculation for atomic nuclei can be determined from the 
sum of the single-particle energies $\sum\epsilon_\mu$ of a suitably defined 
potential.\cite{Strutinsky,Brack72} A similar result holds for $\tilde{E}(N)$ 
extracted from local-density functional calculations.\cite{HarrisArgument} 
The latter are more common for metal clusters.\cite{ekardt84,GenzkenPRL,%
BrackRev} For clarity and to fix the notation we give a short outline of 
the relevant argument.

To find the ground state energy of a system of $N$ electrons using density 
functional theory we use the Kohn-Sham formalism.\cite{HohenbergKohn,KohnSham} 
Starting from some electron density $n_0(r)$ we have to solve the Kohn-Sham 
equations with the potential
\begin{equation}
  V_{KS}(\vec{r}\,) = V_{ext}(\vec{r}\,) + 
     \int d^3r'\,{n_0(\vec{r}\,')\over|\vec{r}-\vec{r}\,'|} + V_{xc}[n_0] .
\end{equation}
Having found the $N$ lowest eigenstates $\psi_\mu(\vec{r}\,)$ with energies 
$\epsilon_\mu$, an estimate of the total energy of the system is given by the 
variational expression
\begin{equation}\label{Etot}
 E[n_0] = E_{kin}[n_0] + E_{Coul}[n_0] + E_{xc}[n_0] ,
\end{equation}
where the kinetic energy is given by
\begin{eqnarray}
  E_{kin}[n_0] &=& \sum_{\mu=1}^{N} \epsilon_\mu 
   - \int d^3r\,d^3r'\;{n_0(\vec{r}\,)\,n(\vec{r}\,)\over|\vec{r}-\vec{r}\,'|}\\
&& - \int d^3r\;V_{xc}[n_0]\,n(\vec{r}\,)
   - \int d^3r\;V_{ext}(\vec{r}\,)\,n(\vec{r}\,) \nonumber,
\end{eqnarray}
and the Coulomb energy is the sum of the Hartree energy, the interaction of the
electron density with the external potential (e.g.\ the potential arising for 
the ion-cores), and the electrostatic self-energy of the ionic cores
\begin{eqnarray}
  E_{Coul}[n_0] &=&  \nonumber
   {1\over2} \int d^3r\,d^3r'\;{n(\vec{r}\,)\,n(\vec{r}\,') \over 
                                                         |\vec{r}-\vec{r}\,'|}\\
&&+ \int d^3r\;V_{ext}(\vec{r}\,)\,n(\vec{r}\,) + E_I .
\end{eqnarray}
The `new' electron density $n(\vec{r}\,)$ in the above expressions is given by
$\sum_\mu|\psi_\mu(\vec{r}\,)|^2$. 

Let us assume that $n_0(\vec{r}\,)$ was chosen close to self-consistency. Then 
$n(\vec{r}\,)$ will not differ too much from $n_0(\vec{r}\,)$
\begin{equation}
  n(\vec{r}\,) = n_0(\vec{r}\,) + \delta n(\vec{r}\,) ,
\end{equation}
and we can expand the expression for the total energy (\ref{Etot}) in powers 
of $\delta n$. Using 
\begin{eqnarray*}
  {1\over2} \int d^3r\,d^3r'\;{n  \,n   \over |\vec{r}-\vec{r}\,'|} &=& 
 -{1\over2} \int d^3r\,d^3r'\;{n_0\,n_0 \over |\vec{r}-\vec{r}\,'|} \\
&& +          \int d^3r\,d^3r'\;{n_0\,n   \over |\vec{r}-\vec{r}\,'|}
   + {\cal O}^2(\delta n)
\end{eqnarray*}
and
\begin{displaymath}
  E_{xc}[n] = E_{xc}[n_0] + \int d^3r\;
      \underbrace{{\delta E_{xc}[n_0]\over \delta n}}_{=V_{xc}[n_0](\vec{r}\,)}
      \,\delta n(\vec{r}\,)
      + {\cal O}^2(\delta n)
\end{displaymath}
we can, to first order in $\delta n$, write the total energy as a functional of 
only the initial electron density $n_0(r)$:
\begin{eqnarray}\label{EHarris}
 E&=&\sum_\mu \varepsilon_\mu - \nonumber {1\over2}
     \int d^3r\,d^3r'\;{n_0(\vec{r}\,)\,n_0(\vec{r}\,') \over 
                                                        |\vec{r}-\vec{r}\,'|} \\
 &&- \int d^3r\;V_{xc}[n_0](\vec{r}\,)\,n_0(\vec{r}\,) + E_{xc}[n_0] + E_I .
\end{eqnarray}

A good choice for $n_0$ is the electron density $n_{TF}$ resulting from an 
extended Thomas-Fermi (ETF) calculation. Since $n_{ETF}(N;r)$ varies smoothly 
as the number of the valence electrons $N$ in the cluster is changed, all terms 
in (\ref{EHarris}), except for the first one, contribute exclusively to the 
smooth part $\bar{E}(N)$ of the total energy. I.e.\ to first order in 
$\delta n$, all electronic shell effects $\tilde{E}(N)$ are contained in the 
sum of the one-particle energies $\sum \epsilon_\mu$. Hence the oscillating 
part $\tilde{E}(N)$ of the total energy can be determined from the spectrum 
of the family $V(N;r)$ of Kohn-Sham potentials which arise from the electron 
density $n_{ETF}(r)$. More generally, the above reasoning holds for all 
{\em families} of electron densities $n_0(N;r)$ that are close to 
self-consistency and {\em smooth in $N$.} 

It is common practice to immediately work with parameterized potentials 
$V(N;r)$. Usually they are chosen to fit experiments or the results of 
self-consistent calculations.\cite{Knight84,nishioka90,clemenger91,Lerme93b} 
Imagining that these potentials arise from a hypothetical family of electron 
densities, the above arguments still apply. The prototype of such a 
phenomenological shell model is the Woods-Saxon potential\cite{Knight84,%
nishioka90,clemenger91}
\begin{equation}\label{WS}
  V(N;r) = {-V_0 \over 1+\exp\big((r-r_s\,N^{1/3})/a\big)}.
\end{equation}
Variants are the Wine-bottle potential\cite{nishioka90} and the Woods-Saxon 
potential with asymmetric surface\cite{Lerme93b}
\begin{equation}\label{WSasymm}
  V(N;r) = {-V_0 \over 
               1+\exp\left(\big(r-(r_s\,N^{1/3}+\Delta R)\big)/f(r) \right)} ,
\end{equation}
where $f(r)$ is an analytical function modeling the potential near the 
cluster surface.

\begin{figure}
\noindent
\begin{minipage}[t]{3.375in}
  \centerline{\epsfxsize=3.37in \epsffile{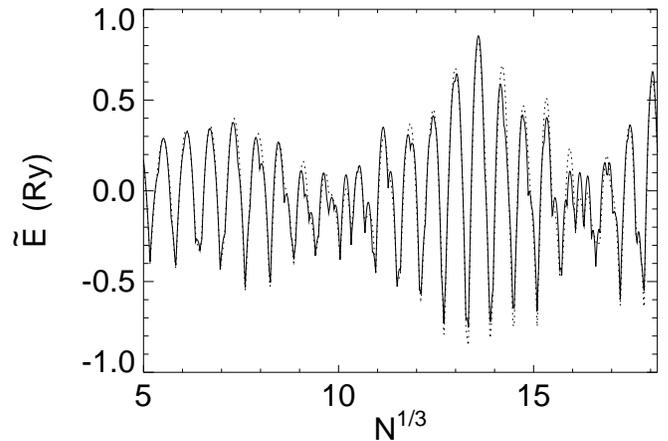}}
  \hspace{2ex}
  \caption[]{\label{Ga_compare} 
     Comparison of the oscillating part $\tilde{E}(N)$ of the total energy 
     obtained from a self-consistent calculation using the homogeneous, 
     spherical jellium model (dotted line) to the $\tilde{E}(N)$ extracted 
     from the sum of the one-particle energies of a family of model potentials
     (full line). The model 
     potentials are given in equation (\ref{WSasymm}), with parameters 
     $r_s=2.19\,a_0$, $V_0=1.04\,Ry$, $\Delta R=0.73 a_0$, $a_0=1.03\,a_0$, 
     $a_1=1.13\,a_0$, and $a_2=0.21\,a_0$.}
\end{minipage}
\end{figure}

To get a feeling for the approximations involved, we compare the results of a 
self-consistent calculation for gallium clusters to the oscillating part of the
total energy found using a family of model potentials (Fig.\ \ref{Ga_compare}). 
For the self-consistent calculations we used the homogeneous, spherical jellium 
model.\cite{ekardt84} The potentials for the one-particle calculation were 
obtained by fitting a function of the type (\ref{WSasymm})
\begin{equation}\label{WSsuper}
  V(N;r) = {-V_0 \over 1+\exp\big((r-R(N))/a(r)\big)} ,
\end{equation}
\begin{figure}
\noindent
\begin{minipage}[t]{3.375in}
  \centerline{\epsfxsize=3.37in \epsffile{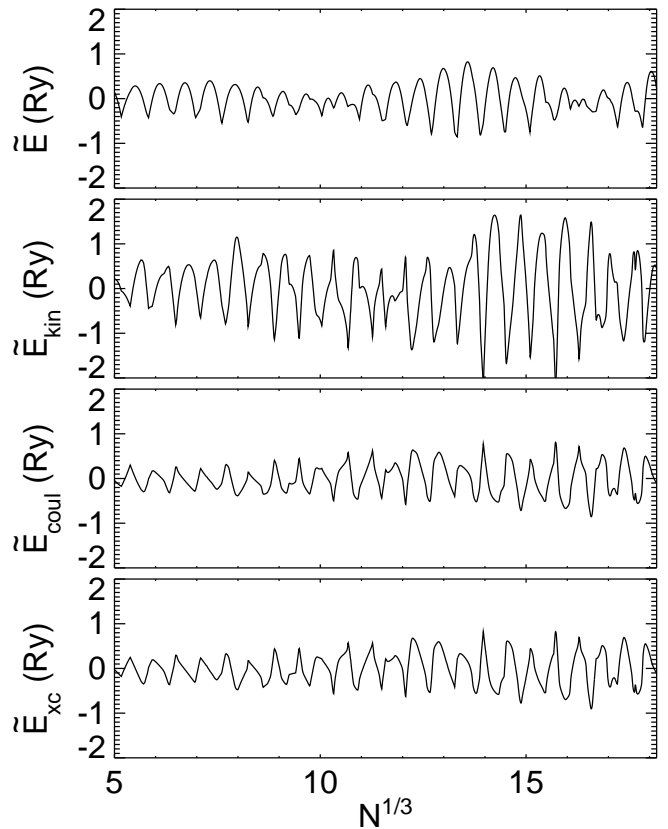}}
  \hspace{2ex}
  \caption[]{\label{Ga_Eosc} 
     Oscillating part of the total energy and of contributions to it 
     (cf.\ eqn.\ (\ref{Etot})), obtained from a self-consistent calculation 
     using the homogeneous, spherical jellium model. 
    }
\end{minipage}
\end{figure}
\noindent
where $R(N)= r_s\,N^{1/3}+\Delta(r)$ and $a(r)= a_0+a_1\tanh(a_2(r-R(N)))$, 
to the self-consistent Kohn-Sham potentials for jellium clusters having 1500, 
3000, 4500, and 6000 valence electrons.

To emphasize the importance of the smoothness of the model potentials $V(N;r)$ 
in $N$, and to demonstrate from what subtle cancellations the electronic shell 
structure arises in self-consistent calculations, we show in Fig.~\ref{Ga_Eosc}
the oscillating part of the total, the kinetic, the Coulomb, and 
the exchange-correlation energy determined from a jellium calculation for 
gallium clusters. Even though in this calculation we are dealing with the 
self-consistent potentials and electron densities (i.e.\ $\delta n =0$),
$E_{kin}(N)$, which contains the sum of the Kohn-Sham energies, is by far 
not the only term contributing to $\tilde{E}(N)$. More surprisingly the 
electronic shell structure as revealed by $\tilde{E}$ cannot be found in 
any {\em single} contributions to the total energy. For the self-consistent 
calculation it rather is resulting from the subtle interplay of the different 
oscillating terms.

\section{Periodic Orbit Expansion (POE)}
\label{POE}

As we have seen in the previous section, the oscillating part $\tilde{E}$ of 
the total energy can be extracted from the sum $\sum\varepsilon_i$ of the $N$ 
lowest eigenenergies for a suitably chosen family $V(N;r)$ of model-potentials. 
The determination of the electronic shell and supershell structure is thus
reduced to an eigenvalue problem for these potentials.
Furthermore, the radius of the clusters we are interested in is considerably
larger than the de Broglie wave-length of the electrons at the Fermi level. 
The semiclassical approximation seems therefore well suited for solving the
single-electron problem in question. In fact, for the spherical 
cavity, a simple rescaling of the Schr\"odinger equation shows that the 
limit $R\to\infty$ is identical to the semiclassical limit $\hbar\to0$.

The salient feature of the semiclassical approach to determining
the electronic shell and supershell structure is, that it provides a
natural splitting of the density of states (and consequently the total
energy) into a smooth and an oscillating part. The smooth part corresponds
to Thomas-Fermi theory, while the quantum corrections are given
by a sum over the nontrivial periodic orbits. For an understanding of the
oscillating part $\tilde{E}$ of the total energy, we need only consider the
latter. 

In the present section we derive the periodic-orbit-expansion (POE) for 
spherical potential wells. Starting from the oscillating part 
$\tilde{\rho}(k)\,dk$ of the density of states for a given potential $V(r)$, 
we proceed to a POE for the oscillating part $\tilde{E}(N)$ of the total
energy for a family $V(N;r)$ of potentials. We illustrate the results by giving
explicit expressions for infinite potential wells. These will be the point of
reference for the leptodermous expansion discussed in Sec.~\ref{lepto}. We
furthermore use the spherical cavity to assess the validity of the various 
approximations made, by comparing the semiclassical results to the results
obtained from numerically solving the Schr\"odinger equation.

\subsection{POE for the density of states}

We first sketch the derivation of the periodic orbit expansion of the density
of states for a spherical potential well along the lines of the path-integral
approach of Gutzwiller.\cite{gutzwiller70,BerryMount,GutzwillerChaos,Brack}
The starting point of the derivation is the relation between the density of
states $\rho(E)$ and the Green's function $G(E)$
\begin{equation}
  \rho(E)\,dE = -{1\over\pi}\Im\,{\rm Tr}\;G(E+i\epsilon)\;dE .
\end{equation}
Expanding the trace in real space leads to 
\begin{equation}\label{trace}
  \rho(E)\,dE = -{1\over\pi}\Im\;\int d^3r\,
              \lim_{\vec{r}\to\vec{r}_0}\,G(\vec{r},\vec{r}_0;E+i\epsilon)\,dE .
\end{equation}
Using the transformation
\begin{equation}\label{trafotE}
  G(\vec{r},\vec{r}_0;E) = 
   -{i\over\hbar}\int_0^\infty dt\,e^{iEt/\hbar}\,G(\vec{r},\vec{r}_0;t)
\end{equation}
the energy-dependent Green's function is expressed in terms of the 
time-dependent Green's function, which in turn can be written as a 
path-integral
\begin{equation}\label{PI}
  G(\vec{r},\vec{r}_0;t) = 
  \int{\cal D}[\vec{r}(\tau)]\;e^{iS[\vec{r}(\tau)]/\hbar} .
\end{equation}
Here the integration is over {\em all} paths $\vec{r}(\tau)$ in 
configuration space connecting the point $\vec{r}_0$ with $\vec{r}$ 
taking a time $t$. $S[\vec{r}(\tau)]$ is the classical action along 
such a path. In the semiclassical limit $\hbar\to0$ only those paths
contribute for which the phase is stationary ($\delta S = 0$: classical
paths). If the second variation of the classical action is finite
($\delta^2 S \ne 0$), the  path-integral becomes a Gaussian integral which
can be evaluated analytically. For $\delta^2 S = 0$ the path-integral 
picks up a phase factor. Thus in the 
semiclassical limit the path-integral (\ref{PI}) decomposes into a sum over 
classical paths\cite{FeynmanHibbs,Schulman}
\begin{eqnarray}
  &&\lim_{\hbar\to0} G(\vec{r},\vec{r}_0;t) = \\
  &&\sum_{\vec{r}_{class}(\tau)}\,           \nonumber
     \sqrt{\partial^2 S\over\partial\vec{r}\partial\vec{r}_0}\;
     \exp\Big(iS[\vec{r}(\tau)]/\hbar + in\pi/2\Big) .
\end{eqnarray}
The transformation (\ref{trafotE}) of the time-dependent into the 
energy-dependent Green's function can also be evaluated by the method of
stationary phase, leading to an expression of the form
\begin{equation}
 G(\vec{r},\vec{r}_0;E) = \sum_{\vec{r}_{class}}\,\sqrt{\Delta}\;
  \exp\Big(iS[\vec{r}] + \phi\Big) .
\end{equation} 
Again there are additional phases associated with the vanishing of the second 
variation.

Taking the trace (\ref{trace}) involves integrating over $\vec{r}_0$ and
taking the limit and $\vec{r}\to\vec{r}_0$. The integration over $\vec{r}_0$ 
is again done in the semiclassical limit. Here the stationary-phase condition 
requires that the final moment equals the initial moment. The limit
$\vec{r}\to\vec{r}_0$ finally closes the orbits. Thus, since the orbits return
to $\vec{r}_0$ with the same momentum, they are closed in phase space, i.e.\ 
they are periodic. There are two distinct classes of such orbits. The first 
consists of only the {\em direct path}, the length of which vanishes as 
$\vec{r}\to\vec{r}_0$. It consequently is {\em local} and gives rise to the 
Thomas-Fermi density of states $\bar{\rho}$. The second class consists of 
periodic orbits of finite length. These {\em non-local} paths give rise to 
a quantum-correction to $\bar{\rho}$. Hence in the semiclassical 
approximation the density of states is given by the local Thomas-Fermi term 
with non-local corrections described by a sum over periodic orbits:
\begin{equation}
  \rho(E)\,dE = \Big(\bar{\rho}(E) + \tilde{\rho}(E)\Big)\,dE .
\end{equation} 

\begin{figure}
\noindent
\begin{minipage}[t]{3.375in}
  \centerline{\epsfxsize=3.37in \epsffile{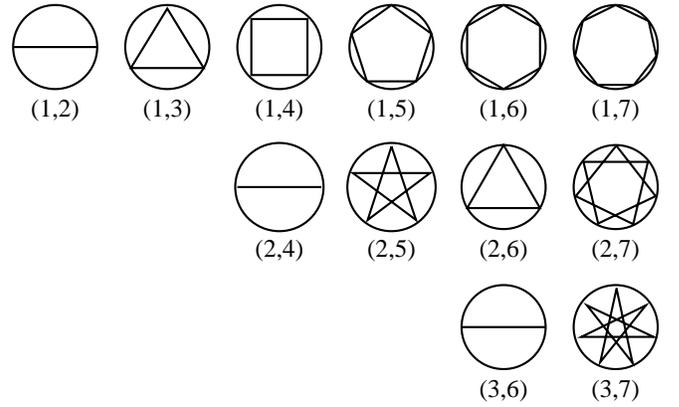}}
  \vspace{2ex}
  \caption[]{\label{orbits} 
   Some periodic orbits for a spherical cavity. They are characterized
   by the pair ${(\lambda,\nu)}$, where $\lambda$ denotes the number of 
   times the orbit revolves 
   around the origin before it closes on itself, and $\nu$ is the number of
   vertices it has. Note that $(n\lambda,n\nu)$ is the orbit obtained by
   traversing ${(\lambda,\nu)}$ $n$ times.}
\end{minipage}
\end{figure}

In a spherical potential well, i.e.\ a potential with at most two radial 
turning points, all periodic orbits can be easily enumerated: A periodic orbit 
is characterized by the number $\lambda$ of times it winds around the
origin and the number $\nu$ of times it traverses the outer turning point.
By symmetry all orbits $(\lambda,\nu)$ that only differ in orientation are
equivalent.
Fig.~\ref{orbits} shows some of the periodic orbits for a spherical cavity. The
periodic orbits for a general spherical potential well are more rounded, but 
are still described by the pairs $(\lambda,\nu)$.\cite{nishioka90,starorb} 

The periodic orbit expansion for the oscillating part of the density of states 
is thus given by a sum over the families of equivalent orbits $(\lambda,\nu)$
\begin{equation} \label{DOS_POE}
  \tilde{\rho}(E)\,dE = \sum_{(\lambda,\nu)} \;A_{(\lambda,\nu)}
   \; \cos\left({S_{(\lambda,\nu)}\over\hbar} 
                 - \varphi_{(\lambda,\nu)}\right) \; dE ,
\end{equation}
where $S_{(\lambda,\nu)}$ is the classical action for an orbit $(\lambda,\nu)$
and $\varphi_{(\lambda,\nu)}$ is the Maslov phase. The amplitude with which the
orbit $(\lambda,\nu)$ contributes is given by
\begin{equation}
  A_{(\lambda,\nu)} = {4\over\sqrt{\pi\nu}}\;{L_{(\lambda,\nu)}\over\hbar}\;
               {\partial s_r/\hbar \over\partial E}
        \left|{\partial^2 s_r\hbar \over \partial L^2} \right|^{-1/2} ,
\end{equation}
with $L$ denoting the angular momentum and $s_r$ the radial 
action.\cite{gutzwiller70}

For a spherical cavity of radius $R_0$ the terms that enter the periodic
orbit expansion take a simple form: The classical action of an orbit equals
its length times the wavevector $k$
\begin{equation}\label{classact}
  S_{(\lambda,\nu)}(k)/\hbar
   =2\nu\,k R_0\;\sin\left({\pi\lambda\over\nu}\right) ,
\end{equation} 

\noindent
the phase is given by 
\begin{equation}
  \varphi_{(\lambda,\nu)} = \left({3\over2}\nu + \lambda - {1\over4}\right)\pi ,
\end{equation}
and the amplitude takes the form 
\begin{equation}
  A_{(\lambda,\nu)} = \sqrt{k}\,R_0^{5/2}\;\alpha_{(\lambda,\nu)} 
\end{equation}
with the dimensionless geometry-factors
\begin{equation}
  \alpha_{(\lambda,\nu)} = {2\over\sqrt{\pi\nu}}\;\;
                 \sqrt{\sin\left({\pi\lambda\over\nu}\right)}\,
                 \sin\left({2\pi\lambda\over\nu}\right) .
\end{equation}
\begin{figure}
\noindent
\begin{minipage}[t]{3.375in}
  \centerline{\epsfxsize=3.37in \epsffile{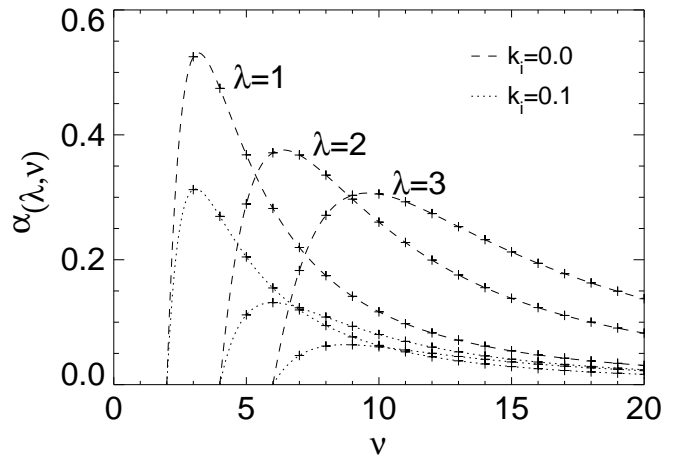}}
  \vspace{2ex}
  \caption[]{\label{alpha}
   Amplitudes $\alpha_{(\lambda,\nu)}$ with which the periodic orbits 
   ${(\lambda,\nu)}$ contribute to the oscillating part $\tilde{\rho}$ of the 
   density of states. Shown are the amplitudes (crosses) for no ($k_i=0$) and 
   for an intermediate ($k_i=0.1$) smoothing. To guide the eye, the amplitudes 
   for a given number of turns are connected by lines.}
\end{minipage}
\end{figure}
The $\alpha_{(\lambda,\nu)}$ determine the relative importance of the periodic
orbits in the POE (\ref{DOS_POE}). Their values for the first few periodic
orbits are shown in Fig.~\ref{alpha} ($k_i=0$). We note that the amplitudes 
for the linear orbits $(\lambda,2\lambda)$ vanish. This can be understood
by a simple dimensional argument.\cite{BaBlo3} Since the sum in the POE is over
{\em all} periodic orbits, the number of different but equivalent
orbits $(\lambda,\nu)$ will be reflected in the amplitude 
$\alpha_{(\lambda,\nu)}$. To parameterize all the different orientations of the
linear orbits it is sufficient to give the coordinates of one of their
outer turning points. Since for a spherical potential well the outer turning 
point lies on the surface of a sphere, the manifold of the linear orbits
has dimension 2. All the higher orbits are not linear but lie in a plane, so
we need an additional parameter to fix the orientation of this plane. The 
manifold of the planar orbits are therefore 3-dimensional (`there are many 
more planar than linear orbits'). Thus the linear orbits do not contribute to
the leading order of the periodic orbit expansion.
The largest amplitudes are
found for the triangular $(1,3)$ and the square $(1,4)$ orbit. The contribution
from other orbits is, however, still large, i.e.\ one has to include many 
periodic orbits in a partial summation of (\ref{DOS_POE}) before one obtains
a result close to $\tilde{\rho}$. This slow convergence is to be expected 
since the density of states for a finite system is given by a sum of 
$\delta$-functions which cannot easily be reproduced by a sum of analytical 
functions. To improve the convergence of the expansion one can replace the 
$\delta$-peaks in the DOS by Lorentzians of width $\gamma$. This corresponds 
to introducing a complex wavevector $k=k_r+ik_i$ in the periodic orbit 
expansion. As can be seen from Fig.~\ref{alpha}, a finite value of $k_i$ serves
to reduce the contribution of higher orbits considerably. However, since the 
shell and supershell structure in metal clusters is not directly linked to 
$\tilde{\rho}$ but rather to the variations $\tilde{E}$ in the total energy, 
we proceed to derive a periodic orbit expansion for $\tilde{E}$. As we will 
see, such an expansion converges much more rapidly than that for the density 
of states. We therefore need not introduce any smoothing. 

\subsection{POE for the total energy}

To find a periodic orbit expansion for the oscillating part $\tilde{E}$ of the
total energy using the POE for $\tilde{\rho}$, we start from the integral 
\begin{equation}\label{Efull}
  E(N)=\int_0^{E_F(N)} E\rho(N;E)\,dE ,
\end{equation}
where the Fermi energy $E_F(N)$ is fixed by the number of electrons $N$ in
the cluster
\begin{equation}\label{Nfull}
  N=\int_0^{E_F(N)} \rho(N;E)\,dE .
\end{equation}
Similar equations hold in Thomas-Fermi theory.
Subtracting the corresponding Thomas-Fermi expression from (\ref{Nfull}) we find
\begin{equation}\label{EF_TF}
  0=\int_{\bar{E}_F(N)}^{E_F(N)} \bar{\rho}(N;E)\,dE
   +\int_0^{E_F(N)} \tilde{\rho}(N;E)\,dE .
\end{equation}
Since the smooth part of the density of states does not vary much over the
small interval $\bar{E}_F\ldots E_F$, we can approximate the first integral
in the above expression by $\tilde{E}_F(N)\,\bar{\rho}(N;E_F)$. We then can
use the above equation to solve for $\tilde{E}_F(N)$.
In a similar fashion we can approximate the difference of (\ref{Efull}) and 
its Thomas-Fermi counterpart by
\begin{displaymath}
  \tilde{E}(N)\approx \tilde{E}_F(N) E_F(N) \bar{\rho}(N;E_F)
                     +\int_0^{E_F(N)} E \tilde{\rho}(N;E)\, dE .
\end{displaymath}
Using the approximate expression for $\tilde{E}_F$ from eqn.~(\ref{EF_TF})
we find
\begin{equation}
  \tilde{E}(N)\approx \int_0^{E_F(N)} (E-E_F(N)) \tilde{\rho}(N;E)\,dE .
\end{equation}
Since the integrand vanishes at the upper limit of integration it is now 
possible to approximate $E_F(N)$ by its Thomas-Fermi counterpart $\bar{E}_F(N)$.
Integrating by parts we finally arrive at
\begin{equation} \label{Eoscint}
 \tilde{E}(N) \approx -\int_0^{\bar{E}_F(N)} dE \int_0^E dE'\,\tilde{\rho}(N;E')
  .
\end{equation}
i.e.\ to find an approximation to the oscillating part of the total energy we
have to integrate twice over the oscillating part of the density of states.
\begin{figure}
\noindent
\begin{minipage}[t]{3.375in}
  \centerline{\epsfxsize=3.37in \epsffile{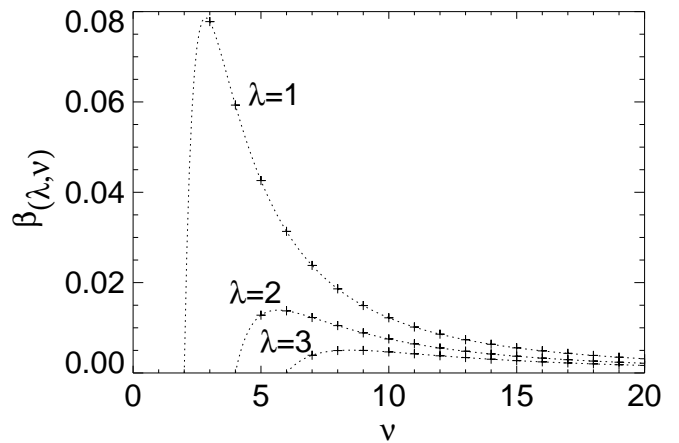}}
  \vspace{2ex}
  \caption[]{\label{beta}
   Amplitudes $\beta_{(\lambda,\nu)}$ with which the periodic orbits 
   ${(\lambda,\nu)}$ contribute to the oscillating part $\tilde{E}$ of the 
   total energy. Comparison with Fig.~\ref{alpha} shows that the dominance of 
   the short, planar orbits, which for the oscillating part $\tilde{\rho}$ of 
   the density of states has to be enforced by introducing an artificial 
   smoothing $k_i$, occurs naturally for $\tilde{E}$ }
\end{minipage}
\end{figure}
\end{multicols}

\noindent
\begin{figure*}
  \centerline{\epsfxsize=6.74in \epsffile{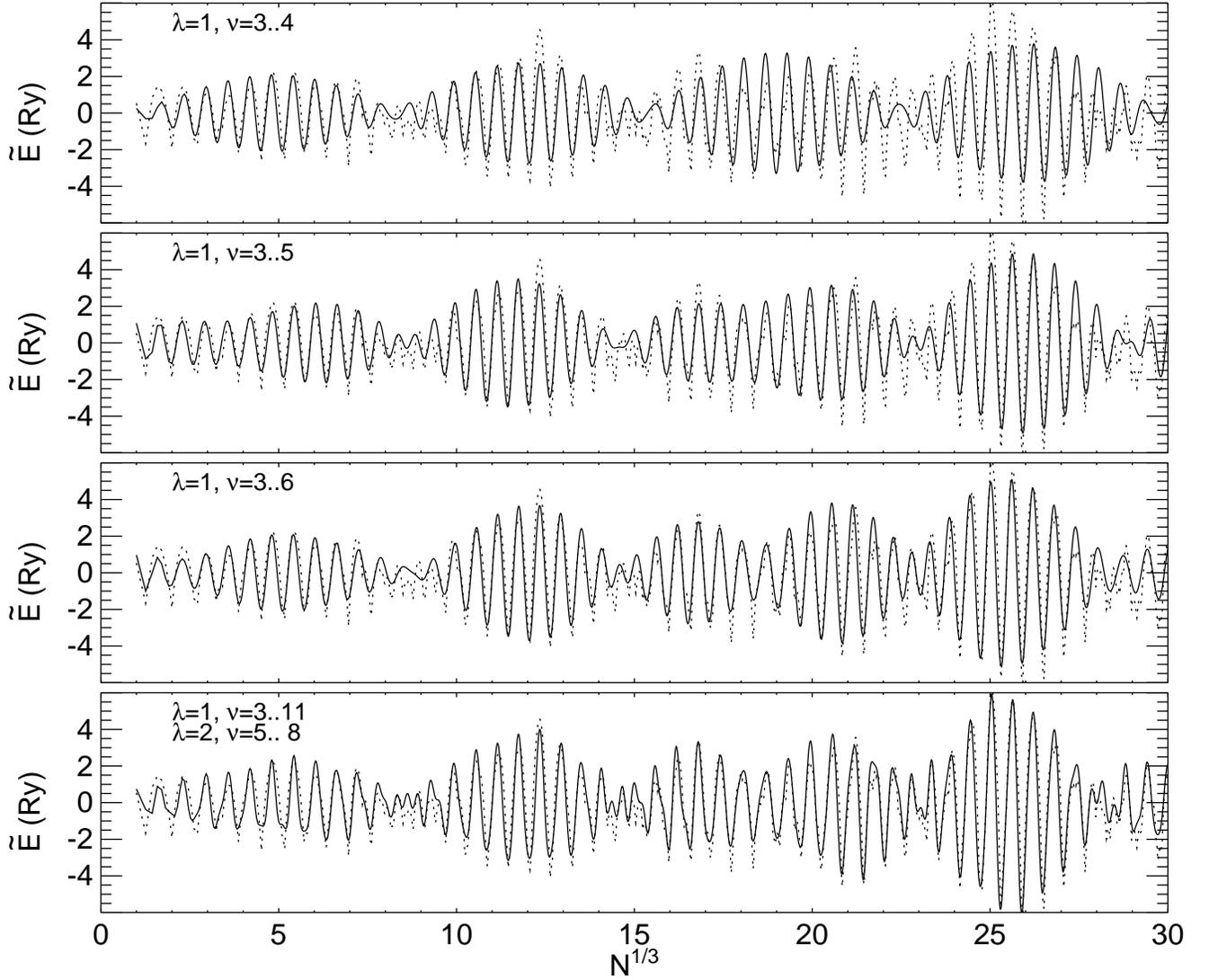}}
  \vspace{2ex}
  \caption[]{\label{caviEosci}
   Comparison of the periodic orbit expansion for the oscillating part 
   $\tilde{E}$ of the total energy to the quantum mechanical result for 
   spherical cavities. The plots show $\tilde{E}$ obtained from a
   truncated periodic orbit expansion, including more and more orbits. The
   orbits were included in the order of decreasing amplitude: $(1,3)$, $(1,4)$,
   $(1,5)$, $(1,6)$, $(1,7)$, $(1,8)$, $(1,9)$, $(2,6)$, $(2,5)$, $(2,7)$, 
   $(1,10)$, $(2,8)$, $(1,11)$. The $\tilde{E}_{QM}$ from the quantum 
   mechanical calculation is shown by the dotted line.}
\end{figure*}

\begin{multicols}{2}
Using (\ref{Eoscint}) and (\ref{DOS_POE}) we find for spherical cavities of 
radius $R_0(N)$ the expansion
\begin{eqnarray} \label{POE2}
  && \tilde{E}(N) \approx \\
  && \sqrt{\bar{k}_F R_0}\,\bar{k}_F^2 \;
     \sum_{(\lambda,\nu)} \, 
       {4\alpha_{(\lambda,\nu)}\over\hat{S}_{(\lambda,\nu)}^2}
       \cos\Big(\hat{S}_{(\lambda,\nu)}\,\bar{k}_F R_0
               - \varphi_{(\lambda,\nu)} \Big) \nonumber
\end{eqnarray}
which is similar to (\ref{DOS_POE}), the main difference being the change in
the amplitudes: Due to the twofold integration the amplitudes are divided 
by the square of the dimensionless classical action $\hat{S}_{(\lambda,\nu)}=
S_{(\lambda,\nu)}/(\hbar k R_0)$. The new geometry factors are thus given by
\begin{equation}
  \beta_{(\lambda,\nu)} := 
    {4\alpha_{(\lambda,\nu)}\over\hat{S}^2_{(\lambda,\nu)}} .
\end{equation}
They are plotted in Fig.~\ref{beta}. A comparison with the 
$\alpha_{(\lambda,\nu)}$ (Fig.~\ref{alpha}) shows how the contributions of the 
long orbits (with large classical action, see eqn.~(\ref{classact})) to the POE
for $\tilde{E}(N)$ are reduced. This improvement of convergence can be 
understood intuitively since $E(N)$ is continuous, while the density of states 
is highly singular, being a forest of $\delta$-functions.

To check the approximations made in the derivation of eqn.~(\ref{Eoscint}) we
compare the results of a truncated periodic orbit expansion for $\tilde{E}$
with the oscillating part $\tilde{E}_{QM}$ of the total energy derived from 
a quantum mechanical calculation. Such a comparison for spherical 
cavities of radius $R_0=N^{1/3}$ is shown in Fig.~\ref{caviEosci}. It turns
out that the truncated POEs reproduce $\tilde{E}_{QM}$ very well, even
if only a few periodic orbits are included. In particular the first two
supershells can be described using only the triangular and square orbit.
However, for even larger sizes $N$ it seems that higher orbits are needed 
to describe the structure in $\tilde{E}_{QM}$. We note that due to its
nature of being a semiclassical result, the periodic orbit expansion for 
$\tilde{E}$ will not converge to $\tilde{E}_{QM}$ but to its semiclassical 
approximation.

\section{Lattice Contraction}
\label{lattcontr}

As a first application of the periodic orbit expansion for the oscillating
part of the total energy we look at the effects of a lattice contraction
on the electronic shells and supershells. Extended X-ray absorption 
fine structure (EXAFS) measurements on small clusters adsorbed on some
substrate indicate that the next-neighbor distance might decrease with
decreasing cluster size.\cite{EXAFS1,EXAFS2} Such a contraction of the
cluster is also suggested by a continuum description of large clusters:
Since for small clusters the surface-to-volume-ratio increases, the 
surface tension, which has the tendency to compress the cluster, will become 
more and more important. Neglecting this finite-size effect, a spherical cluster
of $N$ atoms will have a radius $R_0=r_s N^{1/3}$, where $r_s$ is the 
Wigner-Seitz radius of the bulk material. Acting against the compressibility
$\kappa$, the surface tension $\sigma$ will reduce the cluster radius:
\begin{equation}
  {\Delta R\over R_0} = -{2\kappa\sigma\over3}\,{1\over R_0} .
\end{equation}
Thus we expect the radius of a cluster of $N$ atoms to be given by
\begin{equation}
  R(N)=R_0+\Delta R = r_s N^{1/3}+\Delta R
\end{equation}
with a size-independent contraction $\Delta R$. A rough estimate using 
the compressibilities and surface tensions of the liquid alkali metals 
just above the melting point gives $\Delta R/r_s \approx 0.1$.

To see how the electronic shells and supershells are affected by such a
lattice contraction we first turn to the special case of the spherical
cavity. Inspecting eqn.~(\ref{POE2}) we see that the periods of the 
oscillations in $\tilde{E}$ are determined by the product $\bar{k}_F R$, 
while all other quantities in the argument of the cosine are independent 
of cluster size. $\bar{k}_F R$ should, however, depend only weakly on the 
cluster radius, since we expect a decrease in cluster radius to be compensated 
by an increase in the Fermi energy. For the spherical cavity we actually have
\begin{equation}\label{kF_TF}
  \bar{k}_F R = b_1 N^{1/3} + b_2 + {\cal O}(N^{-1/3}) 
\end{equation}
with $b_1=(9\pi/4)^{1/3}$ and $b_2=3\pi/8$. I.e.\ for spherical cavities
$\bar{k}_F(N) R(N)$ is independent of $R(N)$ and therefore the {\em position} 
of the electronic shells and supershells are {\em independent} of the actual 
cluster radius $R(N)$.

Looking again at eqn.~(\ref{POE2}) we notice that in the overall prefactor
of the periodic orbit expansion we have an isolated factor $k_F^2$. Since
the Fermi energy obviously depends on the cluster size, we expect the 
{\em amplitude} of $\tilde{E}$ to be affected by a lattice contraction. 
For $R(N)=r_s N^{1/3}+\Delta R$ we find from (\ref{kF_TF})
\begin{equation}
  \bar{k}_F(N) r_s = b_1 + \left( b_2 - b_1{\Delta R\over r_s}\right)
                   + {\cal O}(N^{-2/3}) .
\end{equation}
Thus for small cluster sizes the amplitude of the shell oscillations will
increase with increasing lattice contraction.

\begin{figure}
\noindent
\begin{minipage}[t]{3.375in}
  \centerline{\epsfxsize=3.375in\epsffile{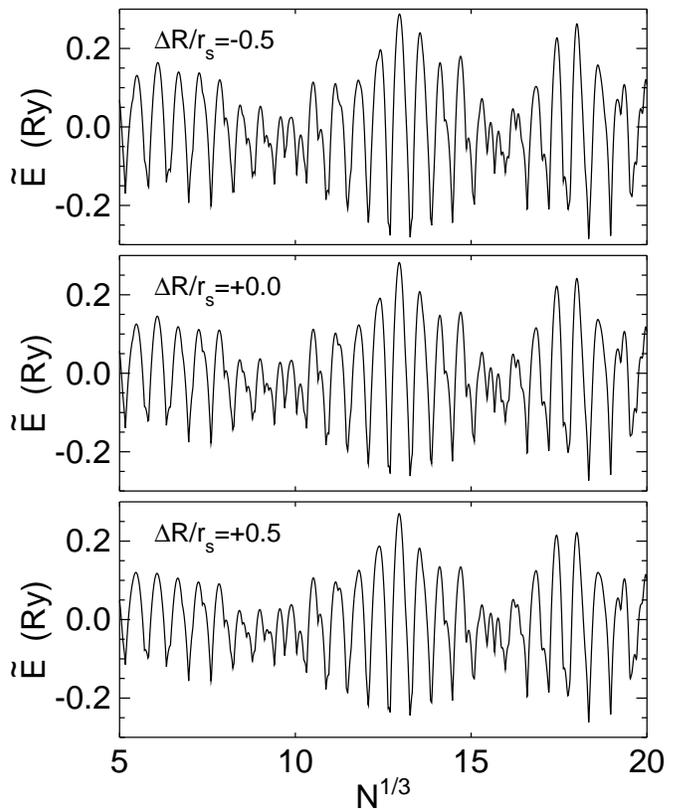}}
  \vspace{2ex}
  \caption[]{\label{WScontr}
   Oscillating part $\tilde{E}$ of the total energy for Woods-Saxon potentials
   $V(r)=-V_0/(1+\exp((r-R(N))/a)$ with lattice contraction 
   $R(N)=r_s N^{1/3} + \Delta R$. The parameters $r_s=3.93\,a_0$, 
   $V_0=0.46\,Ry$, and $a=0.94\,a_0$ were chosen to resemble the potential for
   sodium-jellium clusters. 
  }
\end{minipage}
\end{figure}

To check in how far the above findings also hold for more realistic cluster
potentials, we have calculated the oscillating part of the total energy
from the eigenvalues of a family of Woods-Saxon potentials. The parameters
were chosen to resemble the potentials from jellium calculations for sodium.
In Fig.~\ref{WScontr} we compare $\tilde{E}_{QM}$ for contracted/expanded
clusters with clusters of radius $R(N)=r_s N^{1/3}$. We find that even for
unphysically large contractions/expansions ($\Delta R/r_s=\pm0.5$) the 
location of the electronic shells and supershells is hardly changed, while
for small numbers of atoms $N$ the amplitude of $\tilde{E}(N)$ 
increases/decreases for the contracted/expanded clusters.

\section{Leptodermous expansion}
\label{lepto}

So far we only have explicit expressions of the periodic orbit expansion
for cavity potentials. We now want to extend the POE to more realistic
potentials $V(r)$, like the Woods-Saxon potential (eqn.~(\ref{WS})). 
These potentials differ from the cavity potential by having a surface
of finite width $a$. Since the slope of the cavity potential is infinite 
at the surface, $\delta V(r)=V(r)-V_{cavity}$ is never a small quantity. 
But, rewriting integrals over $V(r)$ in a suitable way, we can use the 
surface width $a$ as an expansion parameter. 

From (\ref{POE2}) we see that we need to find expressions for
(i)   the action integral $\hat{S}_{(\lambda,\nu)}$ for orbits $(\lambda,\nu)$,
(ii)  the Fermi wave-vector $\bar{k}_F$ in Thomas-Fermi approximation, and
(iii) the phases $\varphi_{(\lambda,\nu)}$.  
For this we proceed as follows.
We first introduce the idea of the leptodermous expansion for integrals 
over the potential $V(r)$ for classical action. We find an expansion
\begin{equation}
  \hat{S}_{(\lambda,\nu)}=\hat{S}_{(\lambda,\nu)}^{cavity} 
                         + 2\nu\hat{I}_s\,{a\over R_0} .
\end{equation}
Then we estimate the change in $\bar{k}_F r_s$ due to the finite surface width.
To first order in $a$ we find
\begin{equation}\label{kFexp} 
  \bar{k}_F r_s=\left({9\pi\over4}\right)^{1/3}
               +\left(c_1+c_2{a\over r_s}\right)\,N^{-1/3} .
\end{equation}
Finally we estimate the phase $\varphi_{(\lambda,\nu)}$. Rearranging terms 
in powers of $N^{1/3}$, the argument of the cosine in eqn.\ (\ref{POE2}) 
then reads
\begin{eqnarray}
  &&\left({9\pi\over4}\right)^{1/3}\hat{S}_{(\lambda,\nu)}^{cavity}N^{1/3} \\
  &&+ \hat{S}_{(\lambda,\nu)}^{cavity} \left(c_1+c_2{a\over r_s}\right)
    + 2\nu\,\left({9\pi\over4}\right)^{1/3}\hat{I}_s\,{a\over r_s}
    - \varphi_{(\lambda,\nu)} \nonumber ,
\end{eqnarray}
i.e.\ the first order terms in the leptodermous expansion give rise to a
{\em phase shift} in the periodic orbit expansion, while the frequencies
$S_{(\lambda,\nu)}^{cavity}$ are unchanged.

\subsection{Classical action}
To introduce the basic idea of the leptodermous expansion we first consider 
potentials that differ from the cavity potential only in a small region
around the cluster surface, say for $r>R_0-\alpha$. It is then straightforward 
to split the radial integrals into two parts, one integral over the interior 
$r=0\ldots R_0-\alpha$ and one over the surface region 
$r=R_0-\alpha\ldots r_{out}$.
Thus the radial action can be rewritten as
\begin{eqnarray}\label{Ssplit}
  s_r/\hbar
     &=&\int_{r_{in}}^{r_{out}}    \sqrt{E-V(r) -L^2/r^2}\;dr\\
     &=& \int_{r_{in}}^{R_0-\alpha}  \sqrt{.}\;dr
        +\int_{R_0-\alpha}^{r_{out}} \sqrt{.}\;dr \nonumber .
\end{eqnarray}
The first integral is the action integral for a spherical cavity of radius 
$R_0-\alpha$, which we know already. The second integral can, in general, not 
be solved analytically. But for small $\alpha$ we can get a good approximation
by neglecting the variation of the angular momentum term over the small 
interval $[R_0-\alpha,r_{out}]$, e.g.\ setting $L^2/r^2$ to $L^2/R_0^2$. 

Realistic cluster potentials are not that simple. In a Woods-Saxon potential 
there is no obvious point, that separates bulk from surface. We can still make 
the same ansatz by choosing some small $\alpha$, but we have to make sure that
our result does not depend on our specific choice. To do so, we add and 
subtract the integral $\int_0^{R_0-a} \sqrt{E+V_0-L^2/R_0^2}\,dr$ to 
(\ref{Ssplit}). Using $V(r)\approx-V_0$ for $r<R_0-a$ we find
\begin{eqnarray}\label{Sunsplit}
 s_r/\hbar
  &\approx& \int_{r_{in}}^{R_0-\alpha} \sqrt{E+V_0  - L^2/r^2  }\;dr \\
  &&       +\int_0^{r_{out}}           \sqrt{E-V(r) - L^2/R_0^2}\;dr \nonumber\\
  &&       -\int_0^{R_0-\alpha}        \sqrt{E+V_0  - L^2/R_0^2}\;dr \nonumber
\end{eqnarray}
Using again $L^2/r^2\approx L^2/R_0^2$ for $r\in[R_0-\alpha,r_{out}]$, we can
extend the upper limit of integration of the first and third integral from
$R_0-\alpha$ to $r_{out}$. The first integral is then the radial action for the
cavity potential, the two other terms give the correction due to the soft
surface of the potential $V(r)$.
Introducing dimensionless quantities $\hat{a}=a/R_0$, 
$\hat{s}=s/kR_0$, and $\hat{L}=L/\hbar kR_0$, with $k=\sqrt{E+V_0}$, and
expanding in powers of the reduced surface-width $\hat{a}$, we get
\begin{equation}\label{srexp}
  \hat{s}_r(P,\hat{L},\hat{a}) = \hat{s}_r^{cavity}(\hat{L}) 
            + \hat{I}_s(P,\hat{L})\,\hat{a} + {\cal O}(\hat{a}^2) ,
\end{equation}
where $P=\sqrt{(E+V_0)/V_0}$, and
$\hat{s}_r^{cavity}(\hat{L})=\sqrt{1-\hat{L}^2}-\hat{L}\arccos(\hat{L})$
is the reduced radial action for a spherical cavity. In appendix \ref{linint} 
it is shown how to calculate $\hat{I}_s$ for a Woods-Saxon potential. 

Given the expansion for the radial action, we now proceed to calculate the 
action for a periodic orbit $(\lambda,\nu)$
\begin{equation}\label{totact}
  \hat{S}_{(\lambda,\nu)} 
   = 2\nu\,\hat{s}_r + 2\pi\lambda\;\hat{L}_{(\lambda,\nu)} .
\end{equation} 
The angular momentum $L_{(\lambda,\nu)}$ associated with the orbit can be 
determined from the periodicity condition: In order to close after $\lambda$ 
turns and having traversed the outer turning point $\nu$ times, the angle 
$\Phi$ swept during one radial oscillation must be $\pi\lambda/\nu$. With 
(\ref{srexp}) this leads to
\begin{equation}
  {\pi\lambda\over\nu} = \Phi = -{\partial\hat{s}_r\over\partial\hat{L}}
   = \arccos\left(\hat{L}\right)-{\partial\hat{I}_s\over\partial\hat{L}}\,
     \hat{a} + {\cal O}(\hat{a}^2) .
\end{equation}
Taking the derivative with respect to $\hat{a}$ at $\hat{a}=0$, we can solve
for the first order correction in the reduced angular momentum:
\begin{equation}
  \hat{L}_{(\lambda,\nu)} 
   = \hat{L}_{(\lambda,\nu)}^{cavity} 
   - \sqrt{1-(\hat{L}_{(\lambda,\nu)}^{cavity})^2}\;
     {\partial\hat{I}_s(P,\hat{L}_{(\lambda,\nu)}^{cavity})
      \over\partial\hat{L}}\,\hat{a} .
\end{equation}
We can use this to expand $\hat{s}_r^{cavity}(\hat{L})$ in (\ref{srexp}) 
around $\hat{L}_{(\lambda,\nu)}^{cavity}=cos(\pi\lambda/\nu)$. Inserting into 
(\ref{totact}) we see that the first order correction in $\hat{s}_r^{cavity}$ 
cancels that coming from $\hat{L}$. Thus to first order in $\hat{a}$ the 
reduced action for a periodic orbit $(\lambda,\nu)$ is given by 
\begin{equation}
  \hat{S}(P,\hat{L}_{(\lambda,\nu)},\hat{a})
   = \hat{S}^{cavity}_{(\lambda,\nu)} 
   + 2\nu\,\hat{I}_s(P,\hat{L}_{(\lambda,\nu)}^{cavity})\;\hat{a} 
   + {\cal O}(\hat{a}^2) .
\end{equation}
This result is independent of the specific form of the potential, as long as
an expansion (\ref{srexp}) of the radial action exists.

\subsection{Fermi level}

We now turn to the problem of determining the Fermi wavevector $\bar{k}_F$ 
in the extended Thomas-Fermi approximation for a cluster with $N$ electrons. 
In an infinite system we have $\bar{k}_F r_s=(9\pi/4)^{1/3}$. For a finite 
system there will be corrections arising form the surface
\begin{equation}\label{kFrs_exp}
  \bar{k}_F r_s = (9\pi/4)^{1/3} + c_F\,N^{-1/3} + \cdots
\end{equation}
The surface term can be calculated from the quantum mechanical scattering phase 
$\varphi(k)$ at the surface potential\cite{BaBlo1,surfcorr}
\begin{equation}\label{cF}
 c_F=-{3\over\bar{k}_F^2}
     \int_0^{\bar{k}_F}\left({\pi\over4}-\varphi(k)\right)k dk .
\end{equation}
Assuming that the potential is slowly varying at the outer turning point, we
can determine the scattering phase from the classical action. 

For a slowly varying potential the WKB wave function in the region 
$r\ll R_0-a$ where the potential is practically constant is\cite{Brack}
\begin{equation}\label{u_WKB}
  u_{WKB}(r)\propto \cos\left(\int_r^{r_{out}} k(r)\,dr -{\pi\over4} \right) ,
\end{equation}
while the quantum mechanical wave function is
\begin{equation}
  u_{QM}(r)\propto \cos(k(R_0-r) - \varphi(k)) .
\end{equation}
In the semiclassical limit, i.e.\ for large $R_0$, both expressions should
be equal. Choosing $r=0$, we find
\begin{eqnarray}
  \varphi(k)&=&-\left(\int_0^{r_{out}} k(r)\,dr - k\,R_0 \right) + {\pi\over4}\\
            &=&{\pi\over4} - \hat{I}_s(P,0)\,k\,a + {\cal O}(a^2) ,
\end{eqnarray}
where in the last equation we have used the linearization (\ref{srexp}) of 
the radial action. We note that for $L=0$ the leptodermous expansion is exact,
i.e.\ there are no higher order terms in (\ref{srexp}). From (\ref{cF}) the 
surface parameter thus is
\begin{eqnarray}\label{cF0}
  c_F&=&-{3a\over \bar{k}_F^2}\int_0^{\bar{k}_F}\hat{I}_s(P(k),0)\;k^2 dk \\
     &=&- \left({9\pi\over4}\right)^{1/3} \hat{I}_N(P_F)\,{a\over r_s} 
       \label{cF1} .
\end{eqnarray}
The analytic expression of $\hat{I}_N$ for Woods-Saxon potentials is calculated
in appendix \ref{linint} (eqn.\ (\ref{hatIN})). 

It is interesting to note that we can obtain the same result from 
expanding the Thomas-Fermi integral in powers of $a$ 
\begin{eqnarray}
  N&=&\int_0^{r_{out}}(\bar{E}_F-V(R))^{3/2} r^2 dr \nonumber \\
   &=&{(\bar{k}_F R_0)^3\over3} + (\bar{k}_F R_0)^3\,\hat{I}_N\,\hat{a} 
\end{eqnarray}
and solving for $\bar{k}_F r_s$. In fact, in the approximation considered here,
both approaches are equivalent.

It is important to realize that the above reasoning involves two, possibly
conflicting approximations. In the ansatz (\ref{u_WKB}) for the semiclassical 
wave function we have assumed that the potential is slowly varying, i.e.\ that
$a$ is large enough, while for the leptodermous expansion of the radial action
of the orbits with $L\ne0$ we require that $a$ is small. 

To see how well the expression (\ref{cF0}) for $c_F$ works, we 
compare it to the surface term obtained by fitting the Fermi wavevector 
$k_F(N)$ calculated quantum mechanically for Woods-Saxon potentials holding
$50\ldots8000$ electrons. As expected our approximation approaches the
quantum mechanical result in the limit of large surface width $a$ 
(slowly-varying-potential regime). But it also works quite well for relatively
small $a$. For very small $a$ the approximation of course breaks down, since
our ansatz does not describe the crossover from the slowly-varying-potential 
regime to the potential step at $a=0$, for which the phase in $u_{WKB}(r)$
is $\arctan(\kappa/k)$ instead of $\pi/4$. Using this phase in the above 
derivation, we recover the correct surface parameter for the finite potential 
well.

Since for larger $a$ the approximation to $c_F$ runs roughly parallel to
the true value of the surface term, on might improve the accuracy of the
leptodermous expansion by shifting (\ref{cF0}) by a constant ($a$-independent) 
amount $c_1$. Since the Woods-Saxon potential in one dimension is exactly 
solvable\cite{Landau,Fluegge}, this is straightforward.
\begin{figure}
\noindent
\begin{minipage}[t]{3.375in}
  \centerline{\epsfxsize=3.375in\epsffile{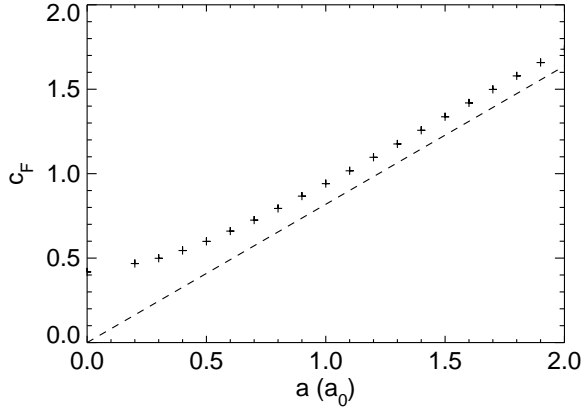}}
  \caption[]{\label{cF_plot} 
    Surface term $c_F$ in the asymptotic expansion of $\bar{k}_F r_s$ 
    (cf.\ eqn.\ (\ref{kFrs_exp})) as a function of the surface width $a$ 
    for a Woods-Saxon potential. The potential parameters  are $V_0=0.45\,Ry$,
    $r_s=4\,a_0$). The crosses were obtained by fitting the Fermi wavevector 
    from full quantum mechanical calculations for clusters with $50\ldots8000$
    electrons. The dashed line gives the estimate of $c_F$ as given by 
    eqn.\ (\ref{cF0}).}
\end{minipage}
\end{figure}

\subsection{Maslov phase}

For separable potentials the Maslov phase is determined by the phases that 
the semiclassical wave function picks up at the classical turning points. 
For a given periodic orbit $(\lambda,\nu)$ there are $2\lambda$ turning points
in the $\vartheta$-motion, each contributing a phase $\pi/2$. The same orbit
also has $2\nu$ radial turning points. The $\nu$ inner turning points see the
smooth centrifugal potential, hence also contribute $\pi/2$. The phase
$\phi_{out}$ at the outer turning point depends on the shape of the potential 
$V(r)$ at the surface. The Maslov phase for the orbit is then given by
\begin{equation}\label{orbitphase}
  \varphi_{(\lambda,\nu)}= [(1/2+\phi_{out}/\pi)\nu + \lambda-1/4]\pi .
\end{equation} 

For a step potential, we can find $\phi_{out}$ by matching the semiclassical 
radial wave function to the boundary condition at the turning point. For an 
infinite potential well $\phi_{out}=\pi$, while for a well of depth $V_0$
$\phi_{out}=2\arctan(\kappa/k)$, where $\kappa=\sqrt{V_0-k^2}$.

For a slowly varying potential, on the other hand, the standard result obtained 
by linearizing the potential around the classical turning point is
$\phi_{out}=\pi/2$.

The Woods-Saxon potential for a typical cluster has a surface width of
$a=0.5\ldots1.5\,a_0$. As we have seen above, for such values of $a$ 
the potential is already in the slowly-varying-regime. We therefore use 
\begin{equation}
  \varphi_{(\lambda,\nu)}= [\nu + \lambda-1/4]\pi .
\end{equation}

\subsection{Leptodermous POE}

We can now collect all the contributions to calculate the effect of a softening
of the potential at the surface on the periodic orbit expansion (\ref{POE2})
of the oscillating part $\tilde{E}$ of the total energy. The frequencies
associated with the periodic orbits turn out to be unchanged, to first order
there is only a phase shift:
\begin{equation}\label{POEshift}
  \tilde{E}(N)\propto\sum_{(\lambda,\nu)} \beta_{(\lambda,\nu)}\cos\left(
   \left({9\pi\over4}\right)^{1/3}\hat{S}_{(\lambda,\nu)}^{cavity}\,N^{1/3} 
    + \Delta\Phi_{(\lambda,\nu)} \right)
\end{equation}
with
\begin{eqnarray}\label{DPhi}
 \Delta\Phi_{(\lambda,\nu)}&=&
  \left({9\pi\over4}\right)^{1/3}
  \left[2\nu\,\hat{I}_s(P,\hat{L})-\hat{S}_{(\lambda,\nu)}^{cavity}\hat{I}_N(P)
  \right]\,{a\over r_s} \nonumber\\
  &&- [\nu + \lambda-1/4]\pi .
\end{eqnarray}

\begin{figure}
\noindent
\begin{minipage}[t]{3.375in}
  \centerline{\epsfxsize=3.375in\epsffile{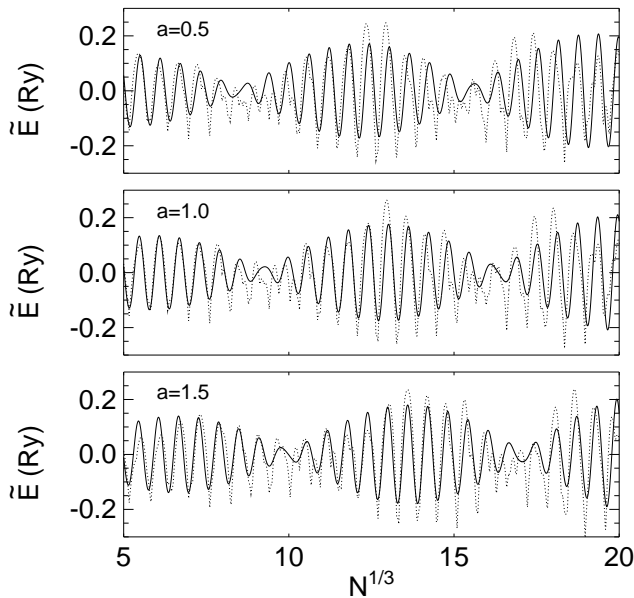}}
  \vspace{2ex}
  \caption[]{\label{shift_plot} 
    Oscillating part $\tilde{E}$ of the total energy for Woods-Saxon potentials
    $V(r)=-V_0/(1+\exp((r-R(N))/a)$ with different surface width $a$. 
    $R(N)=N^{1/3} r_s$ with $r_s=4\,a_0$ and $V_0=0.45$. The dotted lines give 
    the results of quantum mechanical calculations. The full lines are obtained 
    from the leptodermous expansion (\ref{POEshift}), including only triangular
    and square orbits.}
\end{minipage}
\end{figure}

We stress again that we have made two, possibly conflicting approximations.
On the one hand, the leptodermous expansion of the radial action relies on 
the fact that the surface width $a$ is small, while a slowly-varying-potential
assumption enters in the calculation of the scattering phase. To see how the 
above expression works in practical calculations, we compare the periodic orbit
expansion (\ref{POEshift}) with the result of quantum mechanical calculations 
for Woods-Saxon potentials with parameters typical for alkali metal cluster, 
see Fig.\ \ref{shift_plot}. The agreement is surprisingly good. The shift of
the supernodes with increasing surface width is well described, and also
the shell oscillations are quite well reproduced. We could get even better
agreement by numerically fitting the action integrals (see Fig.\ 
\ref{radact_plot}) and the surface coefficient $c_F$ with {\em linear} 
functions in $a$. In that sense the concept of the surface introducing just
a phase-shift in the periodic orbit expansion seems to be applicable even
beyond the range where the analytical expressions from the leptodermous
expansion are good approximations.

Taking only triangular and square orbits into account, the shift in the shell 
(supershell) oscillations is given by 
$(\Delta\Phi_{(1,3)}\pm\Delta\Phi_{(1,4)})/2$. As we can see from (\ref{DPhi}),
the contributions coming from $c_F$ almost cancel for the supershells, since
the classical actions for the triangular and the square orbit are so similar.
Because of this cancellation of errors the shift of the supershells with the 
surface width is very well described in the leptodermous expansion.\cite{dissl}
The shell oscillations are more sensitive to approximations. But the most
important feature, namely that shells are hardly affected by changes in the 
surface width, is also well reproduced. 

\section{Conclusions}

%

Using periodic orbit theory we see that the electronic supershell structure
is a sensitive probe for the surface of clusters. It turns out that possible
lattice contractions, which at first sight seem like important surface effects,
hardly influence the electronic shell structure. There is a pronounced effect 
of the width of the surface region on the position of the supershells. The 
leptodermous expansion around the limiting
case of a spherical cavity provides a natural framework for understanding the
shift of the supershells with increasing surface width. A particularly nice
feature of the leptodermous expansion, as we have presented it here, is the
fact that the expansion of the radial action is the {\em only} input we 
need. All other quantities entering the periodic orbit expansion can be easily 
derived from the radial action. It is therefore straightforward to apply the 
formalism to other types of potentials.

The shift in the electronic supershells that is described by the leptodermous
expansion has been seen in numerical studies of $\tilde{E}$ for soft
potentials\cite{clemenger91} and it has been used to understand the results
of self-consistent jellium calculations.\cite{dissl} The observations of
a shift proportional to the surface width can be regarded as a signature of the
leptodermous regime. Eventually the leptodermous approximation will break
down, since for extremely soft potentials the planar orbits with $\lambda=1$ 
cease to exist. For such potentials star orbits become the leading terms, 
which causes a change in the {\em frequency} of the shell and supershell 
oscillations\cite{Lerme92,Manninen93} as opposed to a change in the {\em phase}
only.

\begin{figure}
\noindent
\begin{minipage}[t]{3.375in}
  \centerline{\epsfxsize=3.3in\epsffile{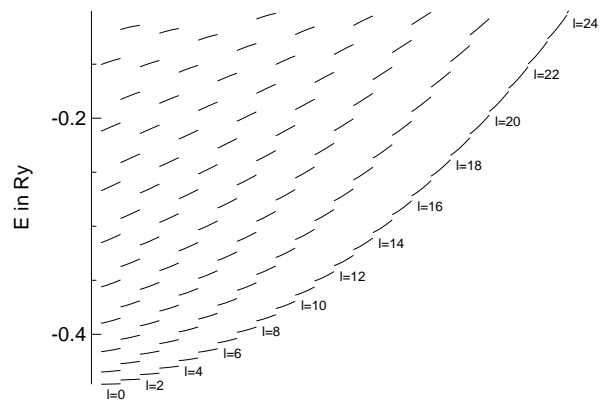}}
  \vspace{2ex}
  \caption[]{\label{shiftlev_plot} 
   Shift in the energy levels $\varepsilon_{n,l}$ for Woods-Saxon potentials
   with different surface-width $a$. The parameters for the potentials are
   $V_0=0.45\,Ry$, $R_0=50\,a_0$ and $a=0.5\ldots1.5\,a_0$. For given angular
   momentum $l$ the levels are plotted one above the other. Each line shows
   $\varepsilon_{n,l}$ as a function of $a$, with $a$ increasing from left to 
   right.}
\end{minipage}
\end{figure}

It is interesting to compare the leptodermous expansion of the semiclassical
sum over periodic orbits with the quantum mechanical perturbation theory. In
quantum mechanics we would expect perturbation theory to break down when 
the shifts of the energy levels are of the order of their spacing. For the 
potentials we have considered here the change in the energy levels is quite 
large, especially for the levels with high angular momentum, which are most 
sensitive to the potential at the surface. Typical shifts of the energy levels 
with the surface width $a$ for a set of Woods-Saxon potentials are shown in 
Fig.\ \ref{shiftlev_plot}. Nevertheless, the electronic shells are not that 
strongly affected, because the levels with large angular momentum, which are 
mostly responsible for the oscillations in the total energy, are shifted by 
large, but similar amounts. It seems that since the semiclassical periodic 
orbit expansion does not deal with individual energy levels but only with the 
collective changes in the spectrum, it works so well, even for large surface 
widths $a$.

The leptodermous expansion should also be of use in understanding experiments 
probing the transport properties of high-mobility semiconductor 
microstructures.\cite{Lindelof} The oscillations in the conductance of quantum 
dots are quite well described by simple cavity potentials. At first sight
this seems surprising since the confining potential of a quantum dot is 
expected to be rather smooth. However, if these potentials are still in the
leptodermous regime, it is clear that calculations using simple cavity 
potentials (or billiards) already describe the qualitatively correct physics.

\section*{Acknowledgments}

It is a pleasure to thank O.~Gunnarsson for his invaluable advice.
Helpful discussions with T.~P.~Martin and M.~Brack are gratefully acknowledged.
A.~Burkhardt, A.~Schuhmacher, and K.~R\"o\ss mann did a great job supporting us
with Computer Algebra Tools.

\begin{appendix}
\section{Leptodermous expansion for a Woods-Saxon Potential}
\label{linint}

In this appendix we show how to evaluate the integrals in the 
leptodermous expansion for a Woods-Saxon potential (\ref{WS}). 

\subsection{Radial action}

For a potential with small surface width $a$ the radial action can be
approximately written as (see discussion
of eqn.\ (\ref{Sunsplit}))
\begin{eqnarray}\label{SrWS}
  s_r/\hbar
     &\approx& \int_{r_{in}}^{R_0} \sqrt{E+V_0  - L^2/r^2  }\;dr \\
     &&       +\int_0^{r_{out}}    \sqrt{E-V(r) - L^2/R_0^2}\;dr \nonumber\\
     &&       -\int_0^{R_0}        \sqrt{E+V_0  - L^2/R_0^2}\;dr \nonumber .
\end{eqnarray}
The first integral is the radial action for the spherical cavity. The third
integral is trivial: the integrand is a constant. Introducing 
$k=\sqrt{E+V_0}$ and $\hat{L}=L/\hbar kR_0$ to simplify the notation, it is 
given by $\hbar k R_0 \sqrt{1-\hat{L}^2}$.
The second integral is more difficult. Rewriting the Woods-Saxon potential 
(\ref{WS}) as
\begin{equation}\label{WSrewrite}
  V(r)=-V_0+{V_0\over2}\left(1+\tanh\left({r-R_0\over2a}\right)\right)
\end{equation}
and substituting $y=(r-R_0)/a$, we are led to
\begin{equation}\label{sinteg0}
  {\sqrt{2}\hat{a}\over P} \int_{-1/2\hat{a}}^{{\rm artanh}(c)} \sqrt{c-\tanh(y)}\;dy
\end{equation}
with $c=2P^2(1-\hat{L}^2)-1$, $P=\sqrt{(E+V_0)/V_0}$, and $\hat{a}=a/R_0$.
This expression can be evaluated analytically
\begin{eqnarray}\label{sinteg}
  &2\hat{a}\,
  \left\{\sqrt{1-\hat{L}^2}\,
  {\rm artanh}\sqrt{1-\hat{L}^2-{1+\tanh({-1/\hat{a}})\over2P^2}
                      \over 1-\hat{L}^2}\right. \\
  &\left.-\sqrt{{1\over P^2}-(1-\hat{L}^2)}\,
  \arctan\sqrt{1-\hat{L}^2-{1+\tanh({-1/2\hat{a}})\over2P^2}
               \over 1/P^2- (1-\hat{L}^2)}\right\} \nonumber .
\end{eqnarray}
Since the ansatz (eqn.\ (\ref{SrWS})) is already a first order approximation,
we need only the expansion of the above expression for small $a$. For the 
second term this is straightforward 
\begin{equation}
   \arctan\sqrt{.}
   =\arcsin\left(P\sqrt{1-\hat{L}^2}\right) 
    + {\cal O}(\hat{a}^2) .
\end{equation}
The first term is a bit more difficult, since for $\hat{a}\to0$ the 
${\rm artanh}\sqrt{.}$ diverges as $1/2\hat{a}$. We find
\begin{equation}
  2\hat{a}\,{\rm artanh}\sqrt{.}
   = 1+2\hat{a}\,\ln\left(2P\sqrt{1-\hat{L}^2}\right) + {\cal O}(\hat{a}^2) .
\end{equation}
Collecting our results we see that we can expand the radial action for a
Woods-Saxon potential with surface parameter $a$ around the radial action of
the corresponding spherical cavity. The term of first order in $a$ is given
by $\hat{I}_s(P,\hat{L})\,\hat{a}$ with
\begin{equation}\label{hatIs}
  \hat{I}_s(P,\hat{L})= {2\over P}\left(P_L\,\ln(2P_L) 
                      - \sqrt{1-P_L^2}\;\arcsin(P_L)\right) ,
\end{equation}
where we have introduced $P_L=P\sqrt{1-\hat{L}^2}$.
\begin{figure}
\noindent
\begin{minipage}[t]{3.375in}
  \centerline{\epsfxsize=3.375in\epsffile{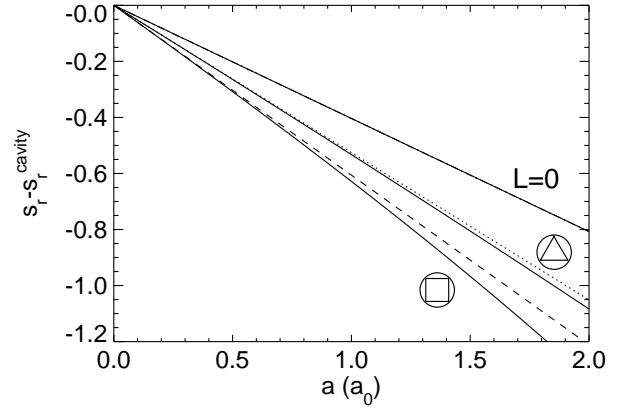}}
  \caption[]{\label{radact_plot}
   Change of the radial action (in units of $\hbar$) for Woods-Saxon 
   potentials with different surface width $a$. The parameters for the 
   potential are $V_0=0.45\,Ry$ and $R_0=40\,a_0$. The full lines give 
   the radial action $s_r(E)$ for a triangular and a square orbit. 
   $k=(9\pi/4)^{1/3}\,1/r_s$. The dotted line shows the change in the 
   radial action for the triangular orbit as calculated using the leptodermous 
   expansion ($\hat{I}_s\;ka$). The dashed line shows the same
   for the square orbit.}
\end{minipage}
\end{figure}

To check how good this linearization of the radial action works in practice,
we compare it to numerical results. Fig.\ \ref{radact_plot} shows such a 
comparison for a Woods-Saxon potential that roughly resembles a sodium cluster
with 1000 electrons. As can be seen, the leptodermous expansion works quite 
well. It is exact for $L=0$ (which is the quantity entering in the calculation
of $c_F$). For the triangular orbit the approximation is very good even for
rather large surface widths $a$. For the square orbit the expansion still works
well, although we start to see deviations for larger $a$. This is due to the 
fact that in the leptodermous expansion we make an approximation in the angular
momentum term, which becomes more critical for orbits with large $\nu$.
It is worth noting that the change in the radial action is surprisingly
linear in $a$. That means, we could obtain results in the spirit of the 
leptodermous expansion (i.e.\ having only phase shifts in the periodic orbit
expansion), by fitting the results of numerical calculations of the action 
integrals with a linear function. Using such fits we could improve the accuracy 
of our results.

\subsection{Fermi level}
\label{TFlin}

From equation (\ref{cF0}) the surface term for $\bar{k}_Fr_s$ in the 
leptodermous expansion is given by
\begin{equation}
  c_F = -{3a\over\bar{k}_F^2}\;\int_0^{\bar{k}_F} \hat{I}_s(P,0)\,k^2\,dk .
\end{equation}
Inserting (\ref{hatIs}) we find
\begin{equation}
  c_F = -{6V_0^{3/2}a\over\bar{k}_F^2}\int\limits_0^{P_F} 
        \left[P^2\ln(2P) - P\sqrt{1-P^2}\arcsin(P)\right]dP ,
\end{equation}
with $P=k/\sqrt{V_0}$. The first integral is straightforward, the second
is easily evaluated by substituting $y=\arcsin(P)$. We thus obtain 
eqn.\ (\ref{cF1}) with
\begin{equation}\label{hatIN}
 \hat{I}_N(P) = 2\left(\ln(2P)+\left[{1\over P^2}-1\right]^{3/2}\arcsin(P)
                       -{1\over P^2}\right) .
\end{equation}

\end{appendix}

\bibliographystyle{prsty_long}

\begin{thebibliography}{10}

\bibitem{cpl91}
T.~P. Martin, S. Bj{\o}rnholm, J. Borggreen, C. Br\'echignac, P. Cahuzac, K.
  Hansen, and J. Pedersen, Chem.\ Phys.\ Lett. {\bf 186},  53  (1991).

\bibitem{nature91}
J. Pedersen, S. Bj{\o}rnholm, J. Borggreen, K. Hansen, T.~P. Martin, and H.~D.
  Rasmussen, Nature (London) {\bf 353},  733  (1991).

\bibitem{Brechignac93}
C. Br\'echignac, P. Cahuzac, F. Carlier, M. de~Frutos, and J.~P. Roux, Phys.\
  Rev.\ B {\bf 47},  2271  (1993).

\bibitem{Broyer93}
M. Pellarin, B. Baguenard, C. Bordas, M. Broyer, J. Lerm\'e, and J.~L. Vialle,
  Phys.\ Rev.\ B {\bf 48},  17645  (1993).

\bibitem{nishioka90}
H. Nishioka, K. Hansen, and B.~R. Mottelson, Phys.\ Rev.\ B {\bf 42},  9377
  (1990).

\bibitem{BaBlo3}
R. Balian and C. Bloch, Ann.\ Phys.\ (N.Y.) {\bf 69},  76  (1972).

\bibitem{gutzwiller70}
M.~C. Gutzwiller, J.\ Math.\ Phys. {\bf 11},  1791  (1970).

\bibitem{Beck}
D.~E. Beck, Solid State Commun {\bf 49},  381  (1984).

\bibitem{ekardt84}
W. Ekardt, Phys.\ Rev.\ B {\bf 29},  1558  (1984).

\bibitem{Lerme93b}
J. Lerm\'e, C. Bordas, M. Pellarin, B. Baguenard, J.~L. Vialle, and M. Broyer,
  Phys.\ Rev.\ B {\bf 48},  12100  (1993).

\bibitem{Lerme94}
J. Lerm\'e, M. Pellarin, B. Baguenard, C. Bordas, J.~L. Vialle, and M. Broyer,
  Phys.\ Rev.\ B {\bf 50},  5558  (1994).

\bibitem{Lerme95a}
J. Lerm\'e, M. Pellarin, J.~L. Vialle, and M. Broyer, Phys.\ Rev.\ B {\bf 52},
  2868  (1995).

\bibitem{Lerme96}
J. Lerm\'e, Phys.\ Rev.\ B {\bf 54}, 14158 (1996)

\bibitem{Lerme95b}
J. Lerm\'e, M. Pellarin, E. Cottancin, B. Baguenard, J.~L. Vialle, and M.
  Broyer, Phys.\ Rev.\ B {\bf 52},  14163  (1995).

\bibitem{jelly}
E. Koch and O. Gunnarsson, Phys.\ Rev.\ B {\bf 54},  5168  (1996).

\bibitem{dissl}
E. Koch, Phys.\ Rev.\ Lett. {\bf 76},  2678  (1996).

\bibitem{Strutinsky}
V.~M. Strutinsky, Nucl.~Phys.~A {\bf 122},  1  (1968).

\bibitem{Brack72}
M. Brack, J. Damgard, A.~S. Jensen, H.~C. Pauli, V.~M. Strutinsky, and C.~Y.
  Wong, Rev.\ Mod.\ Phys. {\bf 44},  320  (1972).

\bibitem{HarrisArgument}
C. Yannouleas and U. Landman, Phys.\ Rev.\ B {\bf 48},  8376  (1993).

\bibitem{bennemannprl}
P. Stampfli and K.~H. Bennemann, Phys.\ Rev.\ Lett. {\bf 69},  3471  (1992).

\bibitem{EXAFS1}
G. Apai, J.~F. Hamilton, J. Stohr, and A. Thompson, Phys.\ Rev.\ Lett. {\bf
  43},  165  (1979).

\bibitem{EXAFS2}
A. Balerna, E. Bernieri, P. Picozzi, A. Reale, S. Santucci, E. Burattini, and
  S. Mobilio, Surf.\ Sci. {\bf 156},  206  (1985).

\bibitem{PerdewLiquidDrop}
J.~P. Perdew, Y. Wang, and E. Engel, Phys.\ Rev.\ Lett. {\bf 66},  508  (1991).

\bibitem{PerdewClust}
E. Engel and J.~P. Perdew, Phys.\ Rev.\ B {\bf 43},  1331  (1991).

\bibitem{GenzkenPRL}
O. Genzken and M. Brack, Phys.\ Rev.\ Lett. {\bf 67},  3286  (1991).

\bibitem{BrackRev}
M. Brack, Rev.\ Mod.\ Phys. {\bf 65},  677  (1993).

\bibitem{HohenbergKohn}
P. Hohenberg and W. Kohn, Phys.\ Rev. {\bf 136},  B864  (1964).

\bibitem{KohnSham}
W. Kohn and L.~J. Sham, Phys.\ Rev. {\bf 140},  A1133  (1965).

\bibitem{Knight84}
W.~D. Knight, K. Clemenger, W.~A. de~Heer, W.~A. Saunders, M.~Y. Chou, and
  M.~L. Cohen, Phys.\ Rev.\ Lett. {\bf 52},  2141  (1984).

\bibitem{clemenger91}
K. Clemenger, Phys.\ Rev.\ B {\bf 44},  12991  (1991).

\bibitem{BerryMount}
M.~V. Berry and K.~E. Mount, Rep.\ Prog.\ Phys. {\bf 35},  315  (1972).

\bibitem{GutzwillerChaos}
M.~C. Gutzwiller, {\em Chaos in Classical and Quantum Mechanics} (Springer, New
  York, 1990).

\bibitem{Brack}
M.~Brack and R.~K.\ Bhaduri, {\em Semiclassical Physics} 
(Addison-Wesley, Reading, 1997).

\bibitem{FeynmanHibbs}
R.~P. Feynman and A.~R. Hibbs, {\em Quantum Mechanics and Path Integrals}
  (McGraw-Hill, New York, 1965).

\bibitem{Schulman}
L.~S. Schulman, {\em Techniques and Applications of Path Integration} (Wiley,
  New York, 1981).

\bibitem{starorb}
For potentials with a very soft surface some orbits $(\lambda,\nu)$ may 
cease to exist.\cite{Lerme92,Manninen93}. Since we are not concerned with 
such extremely soft potentials, we do not consider that case here. 

\bibitem{BaBlo1}
R. Balian and C. Bloch, Ann.\ Phys.\ (N.Y.) {\bf 60}, 401 (1970).

\bibitem{surfcorr}
Ph.~J. Siemens and A. Sobiczewski, Phys.\ Lett.\ B {\bf 41}, 16 (1972).
The equivalence of eqn.\ (3) with the expression given by Balian and
Bloch is seen by an integration by parts. 

\bibitem{Landau}
L.\ D.\ Landau and E.\ M.\ Lifshitz, {\em Quantum Mechanics} 
(Pergamon, New York, 1977).

\bibitem{Fluegge}
S.\ Fl\"ugge, {\em Rechenmethoden der Quantentheorie}
(Springer, Heidelberg, 1990).

\bibitem{Lerme92}
J. Lerm\'e, M. Pellarin, J.~L. Vialle, B. Baguenard, and M. Broyer, Phys.\
  Rev.\ Lett. {\bf 68},  2818  (1992).

\bibitem{Manninen93}
J. Mansikka-aho and M. Manninen, Phys.\ Rev.\ B {\bf 48},  1837  (1993).

\bibitem{Lindelof}
P.~E. Lindelof, P. Hullmann, P. B\o ggild, M. Persson, and S.~M. Reimann, in:
{\em Large Clusters of Atoms and Molecules}, ed.\ by T.~P. Martin (Kluwer,
Dordrecht, 1996), pp. 89.

\end{thebibliography}

\end{multicols}

\end{document}